\documentclass[12pt,preprintnumbers,eqsecnum,nofootinbib]{revtex4}
\usepackage{amsfonts}
\usepackage{mathrsfs}
\usepackage{amssymb}
\usepackage{amsmath}
\usepackage{graphicx}
\usepackage{graphicx}
\usepackage{epstopdf}
\usepackage{pstricks}
\usepackage{slashed}
\usepackage{mathbbol}
\usepackage{color}
\usepackage{refstyle}

\def\mS{m_S}
\def\D{\Delta}
\def\mD{m_{\Delta}}
\def\be{\begin{equation}}
\def\ee{\end{equation}}
\def\bea{\begin{eqnarray}}
\def\eea{\end{eqnarray}}

\begin{document}
\preprint{OSU-HEP-16-07}
\title{Neutrino Masses and Scalar Singlet Dark Matter}

\author{Subhaditya Bhattacharya$^{\blacktriangle}$ \footnote{Email: subhab@iitg.ernet.in} , Sudip Jana$^{\bigstar}$ \footnote{Email: sudip.jana@okstate.edu} and  S. Nandi$^{\bigstar}$ \footnote{Email: s.nandi@okstate.edu}
}

\affiliation{$^\bigstar$Department of Physics and Oklahoma Center for High Energy Physics,
Oklahoma State University, Stillwater, OK 74078-3072, USA \\
$^\blacktriangle$Department of Physics, Indian Institute of Technology Guwahati, Assam-781039, India.\\.
}

\date{\today}

\begin{abstract}

\section*{Abstract}
We propose a simple extension of the Standard Model (SM) which has a viable dark matter (DM) candidate, as well as can explain the generation of tiny neutrino masses. The DM is an electroweak (EW) singlet scalar $S$, odd under an imposed exact $Z_2$ symmetry, interacting to SM through `Higgs-portal' coupling, while all other particles are even under $Z_2$. The model also has an EW isospin $3/2$ scalar, $\Delta$ and a pair of EW isospin vector, $\Sigma$ and $\bar{\Sigma}$, responsible for generating tiny neutrino mass via the effective dimension seven operator. Thanks to the additional interactions with $\Delta$, the scalar singlet DM $S$ survives a large region of parameter space by relic density constraints from WMAP/PLANCK and direct search bounds from updated LUX data. Constraints on the model from Large Hadron Collider (LHC) has also been discussed.  
\end{abstract}


\maketitle

\newpage
\section{Introduction}
\label{sec:intro}

Evidence of Physics beyond the Standard Model (SM) have essentially come from two discoveries: the existence of non-zero neutrino masses and cosmological evidence for the existence of the Dark matter (DM). Although it is not established whether the DM is of astrophysical origin, or of particle physics, many models have been proposed as an extension of SM to accommodate a stable weakly interacting massive particle (WIMP) which can satisfy DM constraints. Similarly, several models of neutrino mass generation have been proposed to satisfy the observed neutrino masses and mixing. However, to bring them under one umbrella seems harder if not impossible \cite{Davoudiasl:2004be, Bhattacharya:2016lts}. The main incentive for addressing such a cause is that both neutrinos and DMs necessarily have to be coupled weakly to the SM. 

In this note we try to address both the issues of neutrino masses and DM together with a minimal possible extension of the SM. The simplest way of accommodating DM is to assume the existence of a singlet scalar ($S$) which is coupled to the SM through Higgs portal coupling. The stability of such a DM is ensured with imposing a $Z_2$ symmetry under which the DM is odd, while the SM is even. The phenomenology of such a case have been discussed in different contexts for the simplicity and predictability \cite{singlet,singlet-FIMP,singlet-Z3,singlet-other}. However, non-observation of DM in direct search experiments is pushing this model under tight constraints. Excepting for Higgs resonance ($m_S=m_H/2$) region, the singlet scalar DM is essentially ruled out from the direct detection constraints \cite{Feng:2014vea, Bhattacharya:2016ysw} to a very large DM mass. 

However, we demonstrate here that scalar singlet DM with Higgs-portal interaction can still survive without incorporating semi-annihilation \cite{semi-annihilation}, or multi-component feature \cite{Bhattacharya:2016ysw} if we can think of additional interactions of such DM to annihilate to some non-SM particles and produce right amount of relic density as observed by WMAP \cite{Hinshaw:2012aka} or PLANCK \cite{Ade:2013zuv}. In such a situation the effective DM-SM portal coupling required to generate correct density is {\it reduced} and so is the direct search cross-section; keeping the DM alive after strong bounds of LUX \cite{Akerib:2013tjd} and XENON100 \cite{Aprile:2012nq}. While the presence of additional annihilation channels for the DM to those beyond the SM particles has already been discussed in literature, for example, in MSSM, we highlight the fact here that such a feature helps alleviating the pressure from non-observation of the DM candidate in direct search experiments. This is first possible attempt to bring such a phenomena correlated to the cause of neutrino mass generation mechanism in one model framework to the best of our knowledge.

While we know that neutrino masses can effectively be generated though the existence of additional EW quadruplet scalar ($\Delta$) \cite{Babu:2009aq}, the singlet scalar $S$ can couple to it and can annihilate to them whenever the mass of the DM is larger than $\Delta$ ($\mS>\mD$). This effectively then reduces annihilation cross-sections for $SS \to SM$ and keeps the model alive in a much larger region of parameter space from direct search experiment. We also point out that constraints coming from neutrino masses do not affect too much to the dark sector. 
 
After the sad demise of the 750 GeV diphoton excess at the Large Hadron Collider (LHC) \cite{ATLAS:2016eeo,CMS:2016crm}, we point out the modified bound on the EW Quadruplet and the DM in absence of any signal beyond SM coming from the LHC experiments. 

The paper is organized as follows. In Section \ref{sec:model}, we discuss the extensions beyond SM to accommodate DM and neutrino masses. In Sec. \ref{sec:dm}, we analyze the constraints on the model from relic density and direct detection searches of the dark matter. Numerical simulations for the LHC signatures of our model is discussed  in Section \ref{sec:collider}. Finally in section \ref{sec:conclusions}, we give our conclusions.

\section{Model and Formalism}
\label{sec:model}

Our model is based on the SM symmetry group $SU(3)_{C}\times SU(2)_{L}\times U(1)_{Y}$ supplemented by an unbroken discrete $Z_2$ symmetry. In the fermion sector, in addition to the SM fermions, we add two vector-like $SU(2)$ triplet leptons, $\Sigma $ and $\bar{\Sigma}$. In the scalar sector, in addition to the usual SM Higgs doublet, $H$, we introduce an isospin $3/2$ scalar, $\Delta$, and an EW singlet, $S$. The Singlet scalar, $S$ is odd under $Z_2$, while all other particles in the model are even under $Z_2$. 

Let us mention that the extra particles $\Delta$, $\Sigma$ and $\bar{\Sigma}$ are introduced to generate tiny neutrino masses via the dimension seven operators \cite{Babu:2009aq,Ghosh:2016lnu}. We will discuss more on this after we introduce the scalar potential. This model can explain neutrino masses for reasonable choice of parameters, and can accommodate both normal and inverted hierarchy for the neutrino masses. The singlet scalar, $S$ along with the $Z_2$ symmetry introduced in this framework, provides with a viable candidate for DM, and is one of the major motivations of this work. The particle contents along with their quantum numbers are shown in the Table~\ref{Table1}. 

\begin{table}[htb]
\begin{center}
\begin{tabular}{|c|c|c|}
\hline 
&$SU(3)_{C} \times SU(2)_{L} \times U(1)_{Y}$\\\hline
\small Matter &${\begin{pmatrix} u \\ d \end{pmatrix}}_L\sim(3,2,\frac{1}{3}), u_R\sim (3,1,\frac{4}{3}), d_R \sim (3,1,-\frac{2}{3})$ \\
&$ {\begin{pmatrix} \nu_e \\ e \end{pmatrix}}_L\sim (1,2,-1), e_R\sim (1,1,-2), \nu_R\sim (1,1,-2)$ \\&$ {\begin{pmatrix} \Sigma^{++} \\ \Sigma^{+} \\ \Sigma^{0} \end{pmatrix}}\sim (1,3,2)$, $ {\begin{pmatrix} \bar{\Sigma}^{0} \\ \bar{\Sigma}^{-} \\ \bar{\Sigma}^{--} \end{pmatrix}}\sim (1,3,-2)$ \\
\hline
\small Gauge & $G^\mu_{a,a=1-8}, A^\mu_{i, i=1-3}, B^\mu$ \\
\hline
\small Higgs & ${\begin{pmatrix} H^{+} \\ H^{0} \end{pmatrix}}\sim(1,2,1)$, ${\begin{pmatrix} \Delta^{+++} \\ \Delta^{++} \\ \Delta^{+} \\\Delta^{0}\end{pmatrix}}\sim(1,4,3)$, $ S\sim(1,1,0)$\\
\hline
\end{tabular}
\\
\end{center}
\caption{Matter, gauge and Higgs contents of the model.}
\label{Table1}
\end{table}

The most general renormalizable  scalar potential  consistent with {scalar spectrum of this} model is given by,

\begin{equation}
\begin{split}
V ( H, \Delta)= - \mu_H^{2}H^{\dagger}H + \mu_\Delta^{2}\Delta^{\dagger}\Delta +   \frac{\lambda_{1}}{2}(H^{\dagger}H)^{2} + \frac{\lambda_{2}}{2}(\Delta^{\dagger}\Delta)^{2} \\+  \lambda_{3}(H^{\dagger}H)(\Delta^{\dagger}\Delta) + \lambda_{4}(H^{\dagger}\tau_{a}H)(\Delta^{\dagger}T_{a}\Delta)  + \lbrace\lambda_{5}H^{3}\Delta^{\star} + h.c. \rbrace ,
\end{split}
\end{equation}
\begin{equation}
\begin{split}
V ( H,\Delta,S) = V ( H, \Delta)+\mu_S^{2}S^{2}  + \frac{\lambda_{7}}{2}S^{4} + \lambda_{8}(H^{\dagger}H)S^{2}  \\+ \lambda_{9}(\Delta^{\dagger}\Delta)S^{2},
\end{split}
\end{equation}
where  $\tau_{a}$ and $T_{a}$  are the generators of $SU(2)$ in the doublet {and} four-plet representations, respectively. 

As was shown in \cite{Babu:2009aq}, even with positive ${\mu_{\Delta}}^2$, due to the $\lambda_5$ term in the potential, and the fields $\Sigma$ and $\bar\Sigma$, the neutral component of $\Delta$ acquires an induced VEV  at the tree level, $v_{\Delta} = - \lambda_5 v_H^3 / M_{\Delta}^2$, where $\langle H \rangle =v_H/\sqrt 2$ is the usual EW VEV. This gives rise to effective dimension seven operator $LLHH(H^{\dagger} H)/ M^{3}$, and generate tiny neutrino masses \cite{Babu:2009aq}. The additional singlet S that we have introduced gets no VEV ($v_s = 0$) to keep the $Z_2$ symmetry intact. Hence, we impose the condition $\mu_S^2>0$.

The mass of the neutral member of the quadruplet is given by  \cite{Babu:2009aq,Ghosh:2016lnu}
\begin{equation}
 M_{\Delta}^{2} = \mu_\Delta^{2} + \lambda_{3}v_{H}^{2} + \frac{3}{4}\lambda_{4}v_{H}^{2},
 \end{equation}
 
The mass splittings between the members of $\Delta$ are given by
\begin{equation}
M_{i}^{2} = M_{\Delta}^{2} - q_{i}\frac{\lambda_{4}}{4}v_{H}^{2},
\label{spectrum}
\end{equation}
where $q_{i}$ is the (non-negative) electric charge of the respective field. The mass
splittings are equally spaced and there are two possible mass orderings. For $\lambda_{4}$ positive,
we have the ordering $M_{\Delta^{+++}} < M_{\Delta^{++}} < M_{\Delta^{+}} < M_{\Delta^{0}}$ and for $\lambda_{4}$ negative, we have the ordering $M_{\Delta^{+++}} > M_{\Delta^{++}} > M_{\Delta^{+}} > M_{\Delta^{0}}$.

Later, it will turn out that that the mass splitting $\Delta M$ plays an important role in the decays, specially that of $\Delta^{++}$. So let us make some comments on the allowed vales of $\Delta M$.  As can be seen from the above equation, this mass splitting is arbitrary depending on the value of $\lambda_4$. however, as shown in \cite{Babu:2009aq}, there is an upper limit of $38$ GeV  on $\Delta M$ coming from the constraint on the $\rho$ parameter. There is also a theoretical lower limit of $1.4$ GeV on $\Delta M $\cite{Babu:2009aq}. In our analysis in this paper, we satisfy both limits.


The gauge singlet scalar $S$ which is odd under a $Z_2$ symmetry provides with a simplest DM candidate which has portal interactions with $\Delta$ (through $\lambda_9$) in addition to the SM Higgs (through $\lambda_8$).

The SM scalar singlet $S$ acquires mass through EW symmetry breaking as 
\be
M_S^2=\mu_S^2+\lambda_8 v_H^2/2
\ee

Note here that $\mu_S^2>0$ implies following inequality

\be
M_S^2>\lambda_8 v_H^2/2, ~{\rm or}~ \lambda_8< \frac{2 M_S^2}{v_H^2}
\ee

Being singlet, the scalar-$S$ does not couple to the SM gauge bosons at tree level. The Yukawa interactions involving $S$ and the SM fermions are also forbidden by the EW as well as $Z_2$ symmetry. 

We now address an important point regarding the stability of the DM particle $S$ due to the added $Z_2$ symmetry. It is well known that a discrete symmetry is vulnerable to Planck scale physics due to anomalies unless it is of gauge origin, and satisfies the discrete anomaly-free conditions \cite{Krauss, Banks-etc, Ibanez}. For example, in minimal supersymmetric Standard Model (MSSM), we introduce the discrete symmetry, matter parity, $P_M$ which is of gauge origin and satisfy the discrete anomaly conditions. This prevents the existence of dimension four baryon and lepton number violating operators in the superpotential and guarantees the stability of stability of proton (the proton decay is still possible through dimension five operator in MSSM, however the decay rate within the limit of proton life time). Following the work of Ibanez and Ross \cite{Ibanez}, if the discrete symmetry is $Z_N$, and $q_i$ are the charges of the fermions of the theory under $Z_N$ should satisfy the following condition:

\bea
\sum_i q_i^3=mN+\eta n\frac{N^3}{8}
\label{eq:anomaly-cond}
\eea 

where $\eta=0,1 ~ \rm{for}~N=\rm{even},~\rm{odd}$ respectively and $m$ and $n$ are integers. In our model, all the SM particles are even under $Z_2$, while the singlet scalar $S$ is odd. Thus, the SM fermions present in the model $(Q,u_L^c,d_L^c, L,e_L^c)$ have $Z_2$ quantum numbers $(1,1,1,1,1)$. We now show that our model satisfy the discrete anomaly-free conditions. For the cubic $Z_2^3$ anomaly, we find
\bea
\sum_i q_i^3=\sum_i (1)^3=15=2m+n
\eea  
using Eq.\ref{eq:anomaly-cond} with $m$ and $n$ being integers. This is easily satisfied with for example, with $m=7$ and $n=1$. For mixed gravitational anomaly, we get 
\bea
\sum_i q_i=\sum_i (1)=15=2p+q
\eea 
where $p,q$ are integers and once again can easily be satisfied with $p=7, ~q=1$. 
For mixed anomaly, for example, $Z_2-SU(2)-SU(2)$, we get
\bea
\sum_{\rm{doublet}} q-i= 4 (1)=4=2r
\eea
where $r$ is an integer and is also easily satisfied. Thus our model satisfies all the anomaly free conditions for the imposed $Z_2$ symmetry to be of gauge origin leading to the stability of the DM.


\section{Dark Matter Analysis}
\label{sec:dm}
Scalar singlet extension of SM to accommodate DM through Higgs portal interaction is under tension as the allowed region of relic density space has been ruled out to a very large DM mass excepting for the Higgs resonance by non-observation in direct search experiment, especially the LUX data \cite{Akerib:2013tjd, Bhattacharya:2016ysw}. Possibilities to evade direct search bound for a DM component is an important question and present day DM research has to answer to that query. Here we present one such phenomena that successfully demonstrates a case for scalar singlet DM which can evade the direct search bound allowing the DM valid through a large region of parameter space.

The scalar singlet $S$ introduced here interacts with the scalar quadruplet $\Delta$ and can annihilate through $SS \to \Delta^0 \Delta^0, \Delta^+ \Delta^-, \Delta^{++} \Delta^{--}, \Delta^{+++} \Delta^{---}$ on top of annihilations to SM particles through Higgs portal interactions. Relic density of the DM in the present universe is obtained by the annihilation cross-section of the DM as

\be
\Omega h^2 =\frac{0.1{\rm pb}}{\langle \sigma v \rangle}
\label{eq:Omega-sigmav}.
\ee

The thermally averaged annihilation cross-section $\langle \sigma v \rangle $ can be written in terms of the usual cross-section as 

\be
\langle \sigma v \rangle_{ab \to cd} =\frac{T g_a g_b}{2 (2\pi)^4 n_a^{eq} n_b^{eq}} \int_{s_0}^{\infty} ds \frac{\lambda(s,m_a^2,m_b^2)}{\sqrt{s}} K_1(\frac{\sqrt{s}}{T}) \sigma_{ab \to cd}.
\label{eq:sigmav}
\ee

Here $g_{a,b}$ corresponds to the degrees of freedom of annihilating particles, $K_1(x)$ is the first Bessel function, $s$ corresponds to the center-of-mass energy available for the process with $s_0=(m_a+m_b)^2$, $T$ is the temperature and $\lambda(a,b,c)=(a^2+b^2+c^2-2ab-2bc-2ca)^{\frac{1}{2}}$. $n_{a,b}^{eq}$ represents the equilibrium distributions of the DMs annihilating, which we assume to be non-relativistic and given by
\be
n_X =  \int \frac{g_X d^3 p}{(2 \pi)^3 2E}  \tilde{f}_X,  \hspace{.4 cm}
n_X^{eq} =  \int \frac{g_X d^3 p}{(2 \pi)^3 2E}  \tilde{f}_X^{EQ},    \hspace{.4 cm}   \tilde{f}_X^{eq}  = \frac{1}{e^{E/T}\pm1}
\label{eq:neq}
\ee

In the following analysis, there are two major contributions to the annihilation cross-section of $S$ as has already been mentioned and can be written as 

\be
\langle \sigma v \rangle=\langle \sigma v \rangle_{SS \to SM}+\langle \sigma v \rangle_{SS \to \D\D}
\ee
The first part of the cross-section is well known and the corresponding annihilation cross-sections to fermions, gauge bosons and the the SM Higgs boson, $h$, and the singlet Higgs boson, $S$ can be written as \cite{Bhattacharya:2016ysw}

\bea
(\sigma v_{rel})_{s s \rightarrow f \overline f}&=&\frac{1}{4\pi s \sqrt s} \frac{N_c\lambda_8^2 m_f^2}{(s-m_h^2)^2+m_h^2 \Gamma_h^2}(s-4m_f^2)^\frac{3}{2}\nonumber\\
{(\sigma v_{rel})}_{s s \rightarrow W^+ W^-}&=&\frac{\lambda_8^2}{8\pi} \frac{s}{(s-m_h^2)^2+m_h^2 \Gamma_h^2}(1+\frac{12m_W^4}{s^2}-\frac{4m_W^2}{s})(1-\frac{4m_W^2}{s})^\frac{1}{2}\nonumber\\
{(\sigma v_{rel})}_{s s \rightarrow Z Z}&=&\frac{\lambda_8^2}{16\pi} \frac{s}{(s-m_h^2)^2+m_h^2 \Gamma_h^2}(1+\frac{12m_Z^4}{s^2}-\frac{4m_Z^2}{s})(1-\frac{4m_Z^2}{s})^\frac{1}{2}\nonumber\\
{(\sigma v_{rel})}_{s s \rightarrow h h}&=&\frac{\lambda_8^2}{16\pi s}[1+\frac{3m_h^2}{(s-m_h^2)}-\frac{4\lambda_8 v^2}{(s-2m_h^2)}]^2 (1-\frac{4m_h^2}{s})^\frac{1}{2}.
\label{eq:annihilationSM}
\eea
Where $N_c=3$ is the color factor for quark and $N_c=1$ for leptons, $m_h=125$ GeV is the Higgs mass and $\Gamma_h$ is Higgs decay width at resonance ($\Gamma_{h\to SM}=4.07$ MeV).
The cross-section to $ss \to \D\D$ can be written as 

\begin{equation}
{(\sigma v_{rel})}_{s s \rightarrow \D \D}=\frac{4 \sqrt{s-4m_{\D}^2}}{8\pi s \sqrt s}\lambda_9^2
\label{eq:annihilationDM}
\end{equation}

Where we assumed that the mass of all the charged and neutral components of $\D$ have the same mass. The upper limit on this mass splitting is 38 GeV \cite{Babu:2009aq}. The factor 4 is essentially indicating four different annihilations $SS \to \Delta^0 \Delta^0, \Delta^+ \Delta^-, \Delta^{++} \Delta^{--}, \Delta^{+++} \Delta^{---}$ which contribute equally in absence of a mass difference between them as we have assumed here.

We have inserted the model in micrOMEGAs \cite{Belanger:2014vza} and scan over the DM parameter space. The relevant parameter space of this model is spanned by the two mass parameters: the DM mass,  $M_s$, the common quadruplet mass, $M_{\D}$, and the two couplings, the Higgs portal coupling $\lambda_8$  and $\lambda_9$, the one connecting DM and quadruplet and are given by the set

\be
\{M_s,~M_{\D}, ~\lambda_8, ~\lambda_9\}
\label{eq:parameters}
\ee

In the following we vary the parameters given in Eq.~\ref{eq:parameters} and find the allowed region of correct relic abundance for the DM, $S$  satisfying WMAP~\cite{Hinshaw:2012aka} constraint ~\footnote{The range we use corresponds to the 
WMAP results; the PLANCK constraints $0.112 \leq \Omega_{\rm DM} h^2 \leq 0.128$~\cite{Ade:2013zuv}, though more stringent, do not lead to 
significant changes in the allowed regions of parameter space.}
\bea
0.094 \leq \Omega_{\rm DM} h^2 \leq 0.128 \,.
\label{eq:wmap.region}
\eea

\begin{figure}[htb!]
$$
 \includegraphics[height=5.5cm]{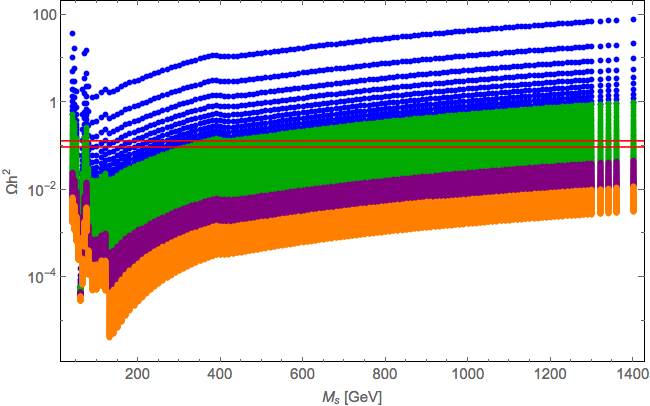}
 \includegraphics[height=5.5cm]{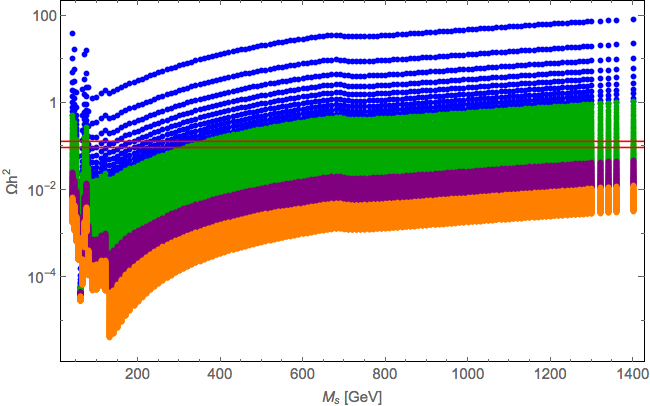}
 $$
 \caption{$\Omega h ^2$ versus $M_s$ for different choices of $\lambda_8=\lambda_9:~ \{0.01-0.1\} ({\rm Blue}),~ \{0.1-0.5\} ({\rm Green}),~ \{0.5-1\} ({\rm Purple}),~\{1-2\} ({\rm Orange})$. Left: $M_{\D}=400~{\rm GeV}$, Right: $M_{\D}=700~{\rm GeV}$ are chosen for illustration. The correct density is indicated through the red horizontal lines.}
\label{fig:Omega-M1}
\end{figure}

In Fig.~\ref{fig:Omega-M1}, we show the variation in relic density of $S$ with respect to the DM mass for fixed value of $M_{\D}=$ 400 GeV on the left and 700 GeV on the right. The couplings $\lambda_8=\lambda_9$ is varied in a long range and indicated through different color codes as follows: $\{0.01-0.1\} ({\rm Blue}),~ \{0.1-0.5\} ({\rm Green}),~ \{0.5-1\} ({\rm Purple}),~\{1-2\} ({\rm Orange})$. The correct density as in Eq.~\ref{eq:wmap.region} is indicated by the red horizontal lines.

\begin{figure}[htb!]
$$
 \includegraphics[height=5.5cm]{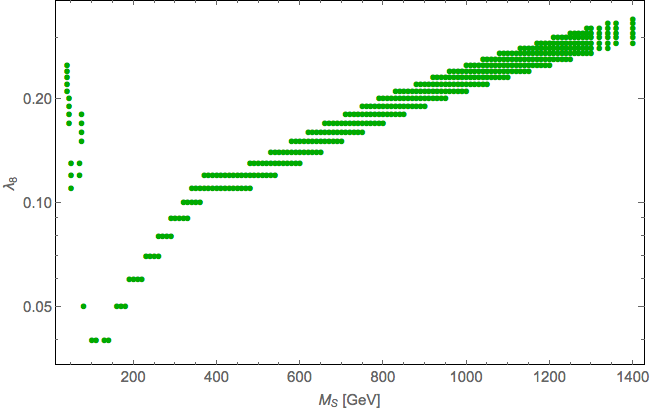}
 \includegraphics[height=5.5cm]{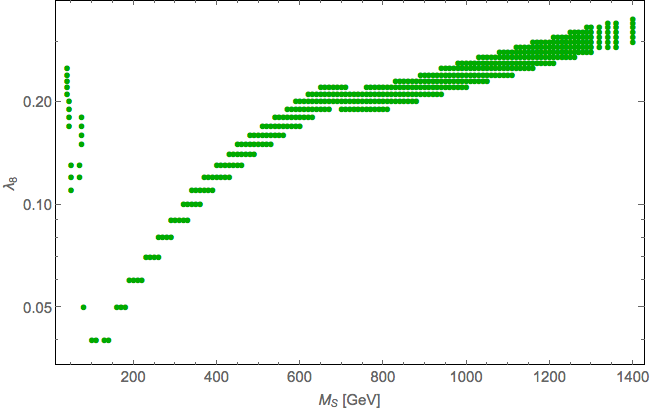}
 $$
\caption{Correct relic density region in $M_s-\lambda_8$ plane. We have assumed $\lambda_8=\lambda_9$ for simplicity here. In the left we have fixed $M_{\D}=400~{\rm GeV}$, on the right: $M_{\D}=700~{\rm GeV}$ chosen for illustration.}
\label{fig:M1-l}
\end{figure}

Next we turn to relic density allowed parameter space of the model. In the simplest scan as shown in Fig.~\ref{fig:Omega-M1}, the allowed region of parameter space can be depicted in $M_s-\lambda_8$ plane with the assumption of $\lambda_8=\lambda_9$ for constant $M_{\D}$. This is shown in Fig.~\ref{fig:M1-l} for fixed values of quadruplet mass as $M_{\D}=400~{\rm GeV}$ on the left and  $M_{\D}=700~{\rm GeV}$ on the right. We have chosen a value so that the chances of conflicting with direct LHC search bound is less. Both in Fig.~\ref{fig:Omega-M1} and Fig.~\ref{fig:M1-l}, we see a clear bump and a subsequent drop in relic density for $M_s>M_\D$ which is 400 GeV on the left and 700 GeV on the right. This is simply due to the additional cross-section of $SS \to \D\D$.

 \begin{figure}[htb!]
$$
 \includegraphics[height=5.5cm]{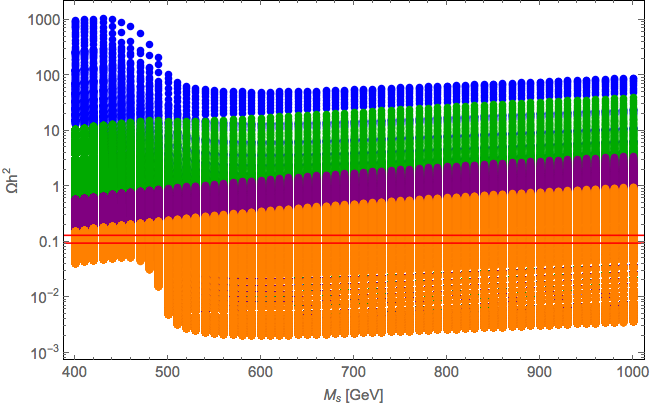}
 $$
 $$
 \includegraphics[height=5.5cm]{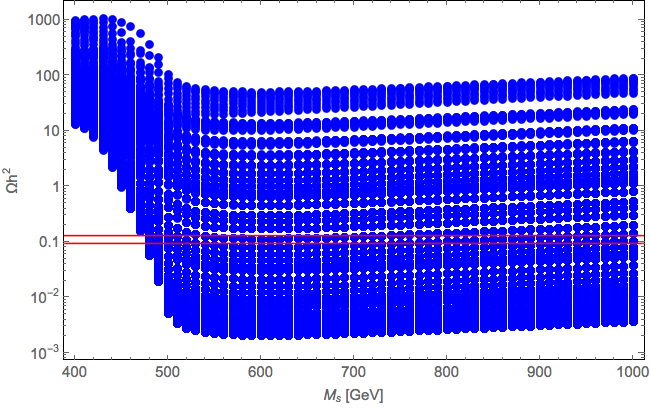}
  \includegraphics[height=5.5cm]{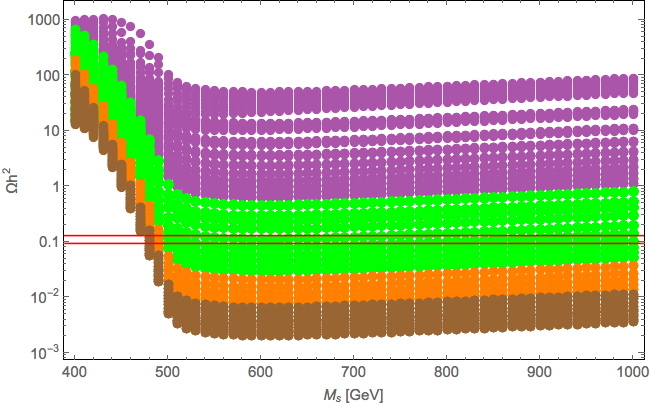}
 $$
 \caption{Top: $\Omega h ^2$ versus $M_s$ for different choices of $\lambda_8:~ \{0.001-0.01\} ({\rm Blue}),~ \{0.01-0.05\} ({\rm Green}),~ \{0.05-0.1\} ({\rm Purple}),~\{0.1-0.2\} ({\rm Orange})$ while $\lambda_9 :\{0.01-2.0\}$ varies. Bottom Left: $\Omega h ^2$ versus $M_s$ for $\lambda_8:~ \{0.001-0.1\} ({\rm Blue})$; Bottom right: $\Omega h ^2$ versus $M_s$ with $\lambda_8:~ \{0.001-0.1\} $ and different choices of $\lambda_9:~ \{0.001-0.1\} ({\rm Purple}),~ \{0.1-0.5\} ({\rm Green}),~ \{0.5-1\} ({\rm Orange}),~\{1-2\} ({\rm Brown})$. $M_{\D}=500~{\rm GeV}$ is chosen for illustration. The correct density is indicated through the red horizontal lines.}
\label{fig:Omega-M2}
\end{figure}

The situation gets even more interesting when we relax the condition imposed on couplings as $\lambda_8=\lambda_9$ and vary them freely as independent parameters. We show such an example in Fig.~\ref{fig:Omega-M2} where we choose the $M_{\D}=500~{\rm GeV}$ for illustration and vary $\lambda_8 :\{0.001-0.3\}$ and $\lambda_9 :\{0.01-2.0\}$ independently. For showing the annihilations $SS \to \D\D$, we highlight the region $M_S \gtrsim M_{\D} $. In the top panel of Fig.~\ref{fig:Omega-M2}, we show $\Omega h ^2$ versus $M_s$ for different choices of $\lambda_8:~ \{0.001-0.01\} ({\rm Blue}),~ \{0.01-0.05\} ({\rm Green}),~ \{0.05-0.1\} ({\rm Purple}),~\{0.1-0.2\} ({\rm Orange})$ while $\lambda_9 :\{0.01-2.0\}$ varies. Each range of chosen $\lambda_8$ actually shows a larger spread as has been pointed out in the bottom left panel, for example with $\lambda_8:~ \{0.001-0.01\}$. Evidently this is due to the large variety of $\lambda_9$ as chosen in the scan. Hence $\Omega h ^2$ versus $M_s$ with $\lambda_8:~ \{0.001-0.1\} $ and different choices of $\lambda_9:~ \{0.001-0.1\} ({\rm Purple}),~ \{0.1-0.5\} ({\rm Green}),~ \{0.5-1\} ({\rm Orange}),~\{1-2\} ({\rm Brown})$ are pointed out in the bottom right panel of Fig.~\ref{fig:Omega-M2}. The correct density is indicated through the red horizontal lines.

\begin{figure}[htb!]
$$
 \includegraphics[height=5.5cm]{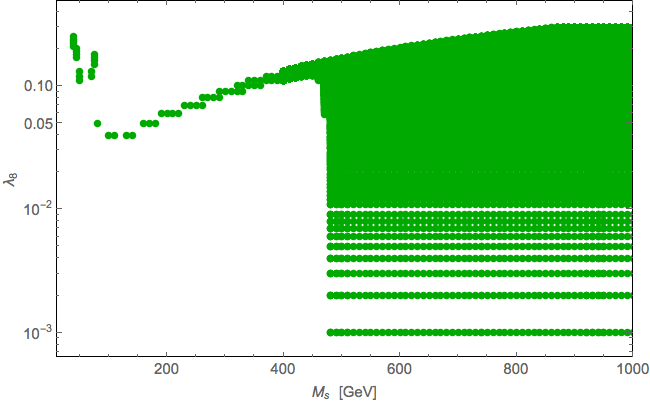}
 \includegraphics[height=5.5cm]{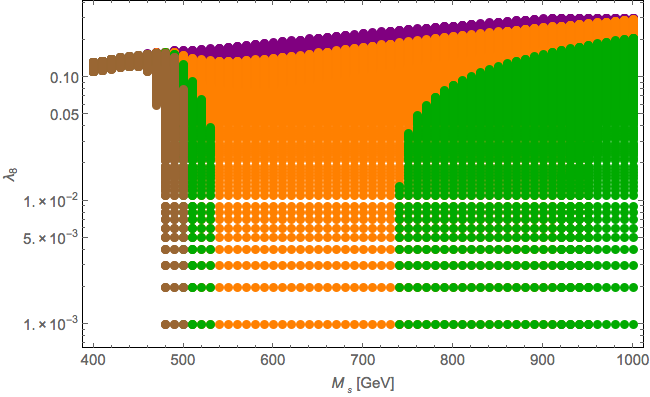}
 $$
\caption{Left: Correct relic density region in $M_s-\lambda_8$ plane when $\lambda_8 :\{0.001-0.3\}$ and $\lambda_9 :\{0.01-2.0\}$ are varied independently. Right: We point out different ranges of $\lambda_9:\{0.1-0.2\} ({\rm Purple}),~\{0.2-0.3\} ({\rm Orange}),~\{0.3-0.4\} ({\rm Green}),~\{0.4-2.0\} ({\rm Brown}) $ in producing correct density in $M_s-\lambda_8$ plane. We choose the $M_{\D}=500~{\rm GeV}$ for illustration and focus in $M_S \gtrsim M_{\D} $.}
\label{fig:M1-l2}
\end{figure}

Now we turn to relic density allowed parameter space in $M_s-\lambda_8$ plane when $\lambda_8 :\{0.001-0.3\}$ and $\lambda_9 :\{0.01-2.0\}$ are varied independently and $M_{\D}=500~{\rm GeV}$ is chosen for illustration as shown in Fig.~\ref{fig:M1-l2}. As is clear from the relic density plot, with a large variation in the $SS -\D\D$ coupling, the allowed plane becomes much larger with even a very small value of $\lambda_8 \sim 10^{-3}$. Different contributions of $\lambda_9$ is shown in the right panel of Fig.~\ref{fig:M1-l2}. The white blanks in between is due to the coarseness of the scanning done in the numerical analysis and do not contain any physics. Similar feature will be observed with any other possible choice of $M_{\D}$. 

\begin{figure}[htb!]
$$
 \includegraphics[height=5.8cm]{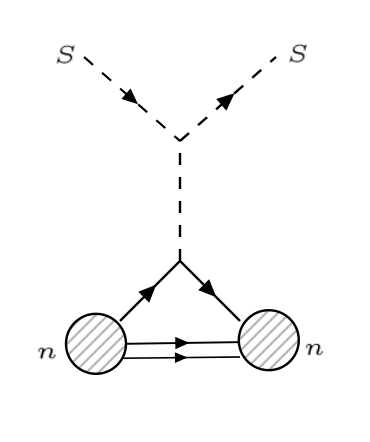}
  $$
 \caption{Feynman diagram for direct detection of the DM.}
 \label{fig:direct}
\end{figure}

Next we turn to direct search constraint for the model. Here the direct detection occurs through Higgs mediation as usual to Higgs portal DM as shown in Fig.~\ref{fig:direct} and the spin independent DM-nucleon cross-section reads:
\bea
\sigma_{SI}^n=\frac{\alpha_n^2 \mu_{n}^2}{4\pi m_{S}^2}
\eea
 where the reduced mass $\mu_n=\frac{m_n m_{S}}{m_n+m_{S}}$ is expressed in terms of  nucleon mass $m_n$, and the nucleon form factors are given by:
  \begin{eqnarray}
  \alpha_n &=& m_n\sum_{u,d,s} f_{T_q}^{(n)} \frac{\alpha_q}{m_q} + \frac{2}{27} f_{T_g}^{(n)} \sum_{q=c,t,b}\frac{\alpha_q}{m_q} \nonumber \\
  &=& m_n\sum_{u,d,s} f_{T_q}^{(n)} \frac{\alpha_q}{m_q} + \frac{2}{27}(1-\sum_{u,d,s} f_{T_q}^{(n)})\sum_{q=c,t,b}\frac{\alpha_q}{m_q} \nonumber \\
  &=& \frac{m_n \lambda_i}{m_h^2}[(f_{T_u}^{(n)}+f_{T_d}^{(n)}+f_{T_s}^{(n)})+\frac{2}{9}(f_{T_u}^{(n)}+f_{T_d}^{(n)}+f_{T_s}^{(n)})]
  \end{eqnarray}
  
Here $n$ stands for both proton and neutron. For proton we choose : $f_{T_u}^p=0.0153$ , $f_{T_d}^p=0.0191$ , $f_{T_s}^p=0.0447$ as the default values in micrOMEGAs. 
 
\begin{figure}[htb!]
$$
 \includegraphics[height=5.5cm]{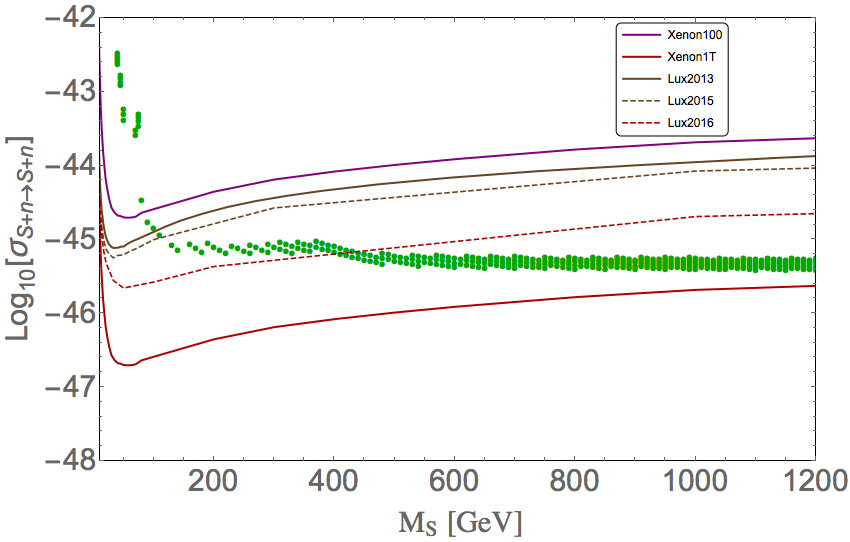}
 \includegraphics[height=5.5cm]{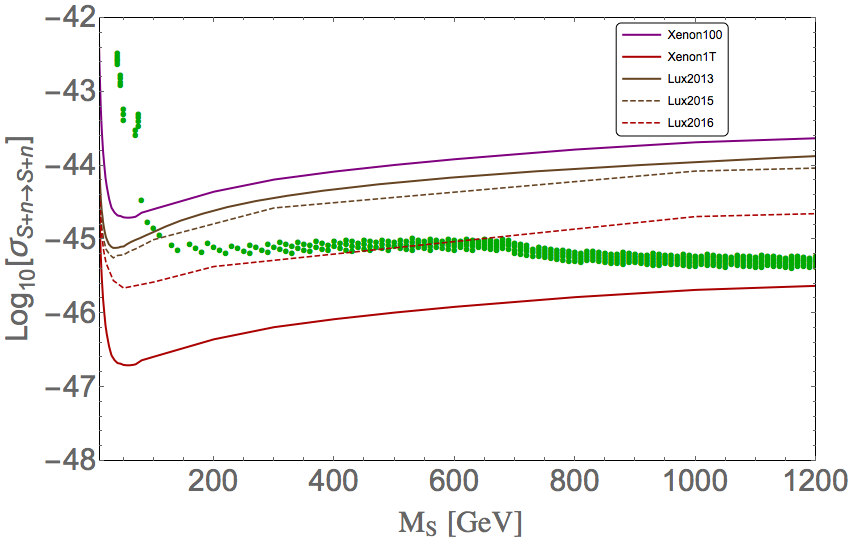}
 $$
 $$
 \includegraphics[height=5.5cm]{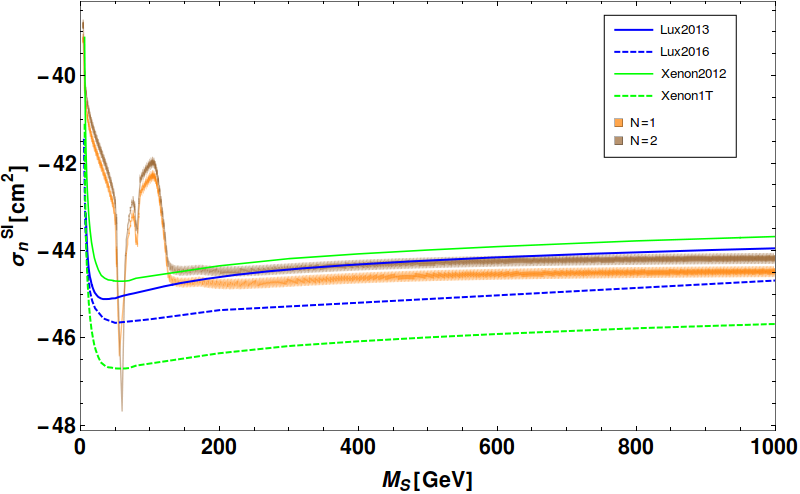}
 $$
\caption{Spin-independent DM-nucleon effective cross-section in terms of DM mass for the case of a fixed $M_{\D}=400$ GeV (top left) and 700 GeV (top right) for points with correct relic density with $\lambda_8=\lambda_9$. XENON100, LUX updated constraints as well as XENON 1T prediction is shown in the figure. Bottom: The case for single component and two-component scalar singlet dark matter with Higgs portal interaction is shown.}
\label{fig:dd1}
\end{figure}

 It is of great importance to see what parameter space of the relic density allowed DM region of the scalar DM model is allowed by the spin independent direct search constraints by XENON100 \cite{Aprile:2012nq}, and LUX \cite{Akerib:2013tjd} data. This is what is presented in Fig.~\ref{fig:dd1} first with the simplified case of $\lambda_8=\lambda_9$.
 
The most important conclusion from the analysis comes out from the direct search results. From Fig.~\ref{fig:dd1} we clearly see that the relic density allowed points of this model is allowed by the direct search constraint from updated LUX data when $M_s>M_{\D}$,  with $M_s$ starting from 400 GeV on the left panel and 700 GeV on the right panel Fig.~\ref{fig:dd1}. This is simply because with additional annihilation cross-section the required coupling to SM drops as shown in Fig~\ref{fig:M1-l} and hence the direct search cross-section also drops to keep the model alive. The smaller the $M_{\D}$, the larger is the allowed region. However, we have to abide by the constraints on the quadruplet mass from collider search bounds and we cant keep it as low as we wish. 

\begin{figure}[htb!]
$$
 \includegraphics[height=6.5cm]{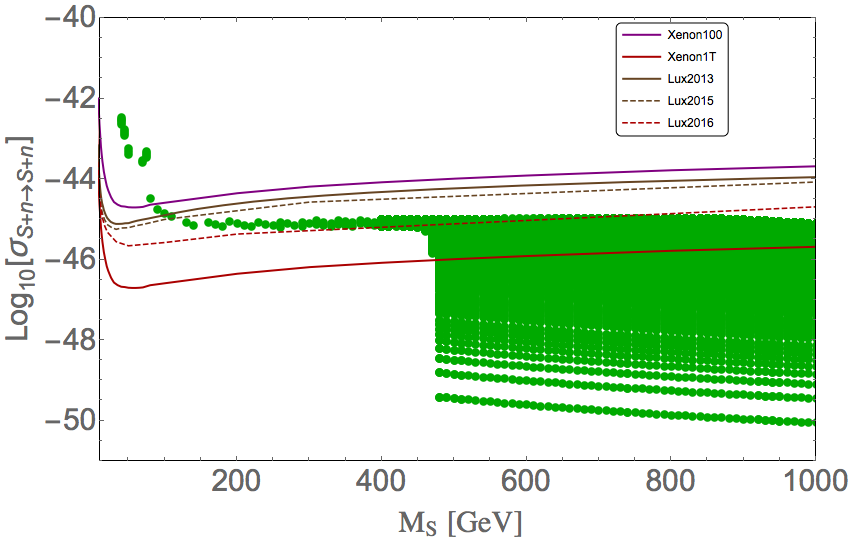}
 $$
 \caption{Spin-independent DM-nucleon effective cross-section in terms of DM mass for the case of a fixed $M_{\D}=500$ GeV when $\lambda_8 :\{0.001-0.3\}$ and $\lambda_9 :\{0.01-2.0\}$ are varied independently for points satisfying relic density. XENON100, LUX updated constraints as well as XENON 1T prediction is shown in the figure.}
\label{fig:dd2}
\end{figure}

The situation is even better for the case when $\lambda_8$ and $\lambda_9$ are varied as uncorrelated parameters as shown in Fig.~\ref{fig:dd2} as illustration for the case with $M_{\D}=500$ GeV. The green points which satisfy relic density for the particular case as indicated in Fig.~\ref{fig:M1-l2}, has a reduced SM Higgs-portal coupling owing to the annihilations to $\D\D$ that do not eventually contribute to direct search. Hence for $M_s>M_{\D}$, as the figure indicate, the direct search may get delayed till XENON1T. This is what makes the model alive for future direct search discoveries. 

In summary the model is allowed by DM constraints in a large region of parameter space with relatively higher values of DM mass, larger than the quadruplet mass.

\section{Constraints from LHC}
\label{sec:collider}
This model provides an interesting avenue to test the neutrino mass generation mechanism at the LHC. The presence of the isospin $3/2$ scalar multiplet $\D$ and a pair of vector-like fermions $\Sigma$ can give rise to rich phenomenology at the LHC. The detailed study of collider signatures has already been studied in early literature \cite{Bambhaniya:2013yca}. From the dark matter perspective, which is main motivation of this work, the most important aspect is the limit on the mass of $\D$ from the latest LHC experimental results.

\begin{figure}[htb!]
$$
 \includegraphics[height=3cm]{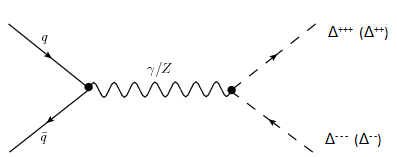}
 \includegraphics[height=3cm]{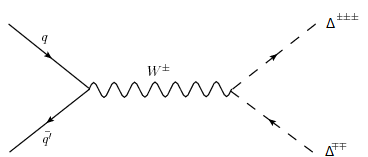}
 $$
 \caption{Feynman diagrams for production of doubly and triply charged scalars at LHC.}
\label{fig:12}
\end{figure}

At the LHC, $\Delta^{\pm\pm\pm} \Delta^{\mp\mp\mp}$, $\Delta^{\pm\pm} \Delta^{\mp\mp}$ and $\Delta^{\pm} \Delta^{\mp}$ are pair produced via the s-channel $\gamma$ and Z exchanges, while $\Delta^{\pm\pm\pm} \Delta^{\mp\mp}$, $\Delta^{\pm\pm} \Delta^{\mp}$ and $\Delta^{\pm} \Delta^{0}$ are pair produced via s-channel $W^{\pm}$ exchange as shown in the Fig.~\ref{fig:12}. 

\begin{figure}[htb!]
$$
 \includegraphics[height=8cm]{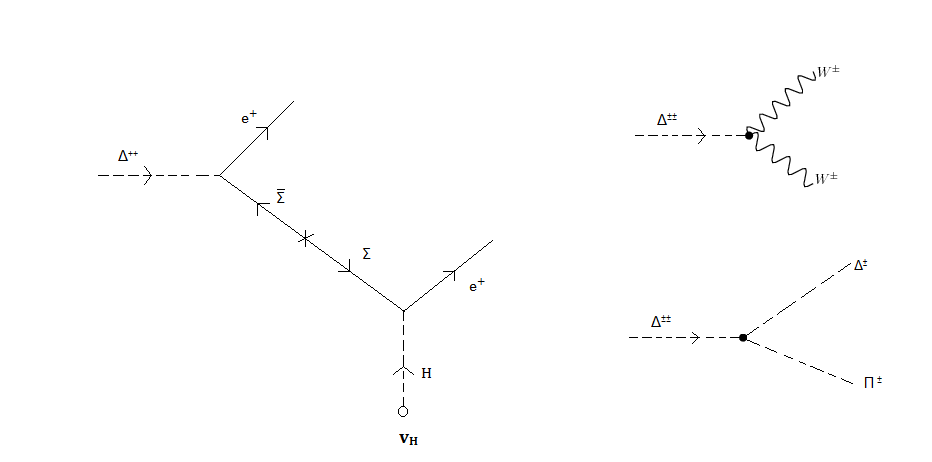}
 $$
 \caption{Feynman diagrams for decay of doubly- charged scalars. }
\label{fig:34}
\end{figure}

\begin{figure}[htb!]
$$
 \includegraphics[height=7cm]{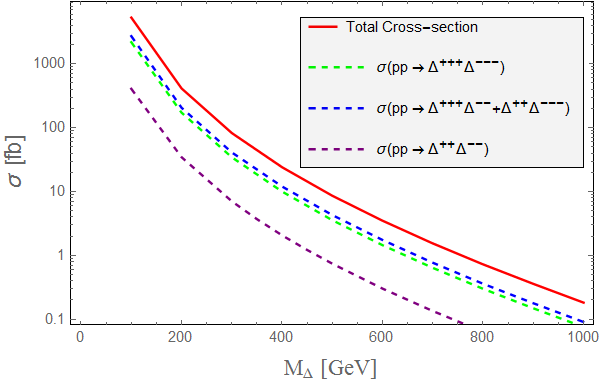}
 $$
\caption{Production cross-sections for triply and doubly charged scalars at the LHC ($\sqrt{S}$ = 13 TeV) as a function of  $M_{\Delta}$.}
\label{fig:5}
\end{figure}

The production cross section of the doubly- and triply- charged scalars at the LHC for center-of-mass energy $\sqrt{s}$ = 13 TeV is shown in Fig.~\ref{fig:5} as a function of the mass parameter. For simplicity, we have taken the masses of the quadruplets to be the same \footnote{Constraints from the $\rho$ parameter dictates the splitting to be $< 38 ~{\rm GeV}$, and can be even smaller depending on the values of $\lambda_4$}.

The model has been implemented in CalcHEP package \cite{Belyaev:2012qa}. For the production cross-section, we used parton distribution function CT10 \cite{Dulat:2013kqa} from LHAPDF library \cite{Whalley:2005nh} with the renormalization and factorization scales being chosen to be the invariant mass of the constituent sub-process.

\begin{figure}[htb]
$$
 \includegraphics[height=4.5cm]{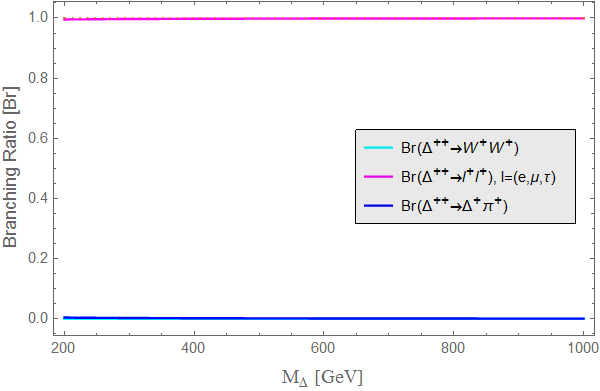}
 \includegraphics[height=4.5cm]{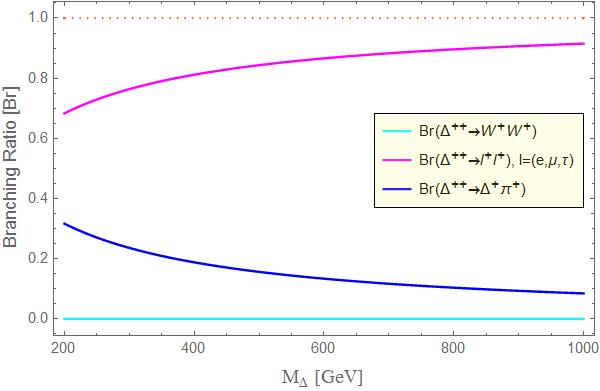}
 $$
 $$
 \includegraphics[height=4.5cm]{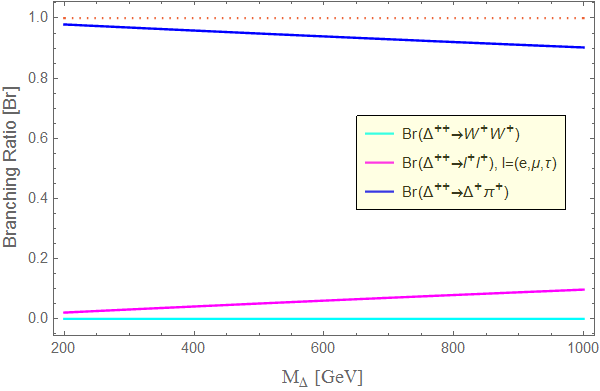}
 \includegraphics[height=4.5cm]{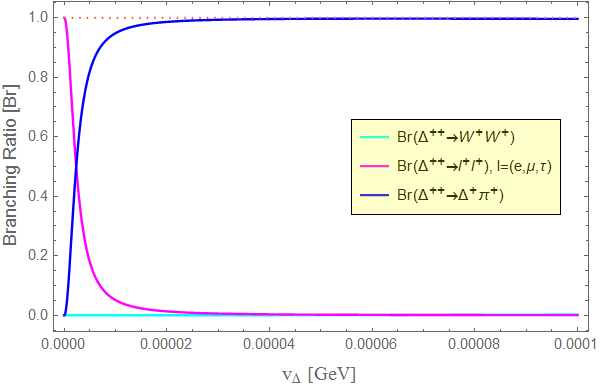}
 $$
 \caption{ Branching ratio (BR) for different decay modes of $\Delta^{\pm\pm}$ as a function of $M_{\Delta}$ for three different values of $v_{\Delta}$ : $10^{-7}$ GeV (Top Left), $10^{-6}$ GeV (Top Right) and $10^{-5}$ GeV (Bottom Left). Bottom Right: Branching ratio (BR) for different decay modes of $\Delta^{\pm\pm}$ as a function of $v_{\Delta}$ considering neutral scalar mass 500 GeV and mass difference between two successive scalars $\Delta M=$ 1.6 GeV.} 
\label{fig:66}
\end{figure}

As mentioned before, our main objective in this collider analysis is to check what constraint the LHC experiments put on the allowed mass of $\D$ which satisfy the dark matter constraints. In our model, $\Delta^{\pm\pm}$, which is a doubly charged scalar,  has three major decay modes : $\Delta^{\pm\pm} \rightarrow l^{\pm}l^{\pm}, W^{\pm}W^{\pm}, \Delta^{\pm}\pi^{\pm}$. All of these modes can give rise same sign dilepton in the final state for $\Delta^{\pm\pm}$ production at LHC. \footnote {A fourth decay mode, to $\Delta^{\pm}W^{{\star}^{\pm}}$, or $\Delta^{\pm\pm\pm} W^{{\star}^{\pm}}$ depending on whether the $\Delta^{\pm\pm\pm}$ is the heaviest or lightest in the quadruplet is possible, but its width is much smaller compared to the other three.}

The ATLAS Collaboration has recently searched \cite{ATLAS:2016pbt} for the doubly-charged Higgs boson in the same-sign di-electrons invariant mass spectrum with luminosity 13.9 $fb^{-1}$ at $\sqrt{S}$ = 13 TeV. Their observed lower mass limit for this doubly charged Higgs, assuming a 100$\%$ branching ratio to di-electrons, is 570 GeV, while the observed lower mass limit, assuming a 50$\%$ branching ratio to di-electrons, is 530 GeV. Our model also must comply with non-observation of excess in same sign dilepton search.

Hence, it is important to parametrise the branching fraction of the doubly charged scalar to be able to compute the same-sign dilepton final state signature arising from the pair production of $\Delta^{\pm\pm}$ at LHC. We note that the decay width for the decay mode $\Delta^{\pm\pm} \rightarrow l^{\pm}l^{\pm}$ is proportional to 1/$v_{\Delta}^{2}$, the decay width to $W^{\pm}W^{\pm}$ final state is proportional to $v_{\Delta}^{2}$, while the one to $\Delta^{\pm}\pi^{\pm}$ is independent of $v_{\Delta}$, and proportional to $(\Delta M)^{3}$. In  Fig.~\ref{fig:66}, we plot the relative branching ratios of $\Delta^{\pm\pm}$ as a function of $M_{\Delta}$ and $v_{\Delta}$. As expected, for a very small 
$v_{\Delta}$, the decay to $l^{\pm}l^{\pm}$ dominate, whereas for higher values of $v_{\Delta}$, the mode $\Delta^{\pm}\pi^{\pm}$ dominate.

\begin{figure}[htb!]
$$
 \includegraphics[height=6.8cm]{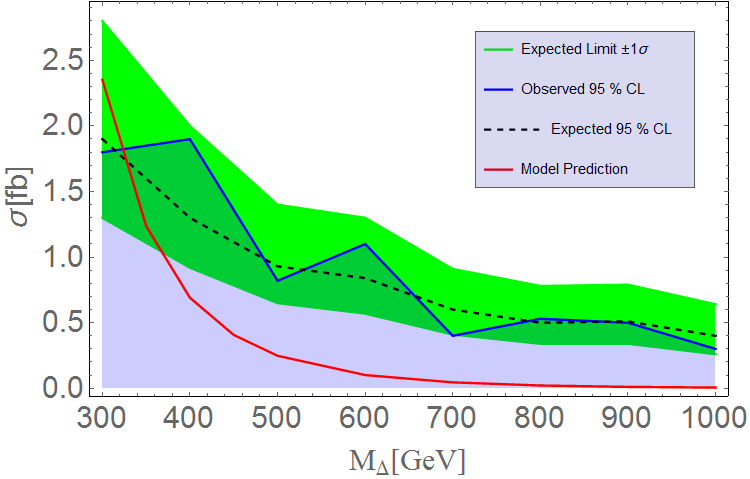}
 $$
 \caption{The model predicted, observed and expected 95$\%$ C.L. upper limits of the cross-section $\sigma$ (pp $\rightarrow \Delta^{\pm\pm} \rightarrow e^{\pm}e^{\pm})$ as a function of $\Delta^{\pm\pm}$ mass at $\sqrt{S}$ = 13 TeV. The limit is derived under the assumption that Br($\Delta^{\pm\pm} \rightarrow e^{\pm}e^{\pm}$) = 33.33$\%$. The red line under the light blue shaded region indicates the allowed mass region.}
\label{fig:67}
\end{figure}

We use the dedicated search by the ATLAS Collaboration \cite{ATLAS:2016pbt} with luminosity 13.9 $fb^{-1}$ at $\sqrt{S}$ = 13 TeV, for the doubly charged scalar  di-electron resonance. Here, to put the most conservative bound on mass $M_{\Delta}$, we assume small value of $v_{\Delta}$ ($\approx 10^{-7}$ GeV) so that $\Delta^{\pm\pm}$ mostly decays into $l^{\pm}l^{\pm}$ with branching ratio nearly equal to one. From our  calculated productions cross sections for the doubly charged scalar, and the branching ratios as shown in Fig~\ref{fig:67}, we obtain a most conservative lower limit of $324$ GeV for the mass of $\Delta^{\pm\pm}$, assuming a 100$\%$ (33.33$\%$) branching ratio to $l^{\pm}l^{\pm}$($e^{\pm}e^{\pm}$). Thus a large mass range of  $\Delta$  satisfying the dark matter constraints are allowed by the latest LHC experimental search. We also note here, that with the assumption of $M_S> M_\Delta$, the additional charged scalars do not decay to DM and hence the branching fractions do not get altered. 

\section{Summary and Conclusions}
\label{sec:conclusions}

In this work, we have analysed the possibility of having a viable DM as a EW scalar singlet $S$, that connects to SM via Higgs portal coupling in a framework that also cater to neutrino mass generation through the presence of a EW quadruplet, $\Delta$, and two EW triplet leptons $\Sigma$ and $\bar{\Sigma}$. An unbroken $Z_2$, under which $S$ is odd, makes the DM stable, while the rest are even. The DM phenomenology is shown to be crucially dictated by the interaction of $S$ with $\Delta$ on top of the Higgs portal coupling that it posses. This is because of the additional annihilation channels for the DM, $S$ to $\Delta$'s on top of SM particles whenever $M_S > M_\D$. In order to satisfy relic density constraints, the coupling of $S$ to the SM particles (Higgs-portal) is reduced compared to the case when annihilations only to SM has to contribute to the whole DM relic density. As a result, direct detection cross sections obtained for $M_S > M_\D$ satisfy the current LUX 2015 data for a large range of the DM parameter space. There is some possibility that the level of direct detection cross section predicted by the model might be within the reach of the XENON1T experiment, while it may go beyond as well. 

This phenomena can be generalised by postulating whenever there is an additional channel for the DM to annihilate, but that doesn't give a contribution to direct search experiments for that DM, one can satisfy the stringent direct search bounds coming from non-observations of DM in terrestrial experiments. The particular example we have presented, has an additional motivation of explaining neutrino masses and the additional DM interaction naturally fits into the model framework with minimal assumptions. 

The validity of the DM parameter space crucially depends on the mass required for the quadruplet $\Delta$. Therefore we attempted to evaluate the current bound on the mass of $\D$ from the latest LHC data. This methodology depends on the non-observation of a doubly charge scalar resonance decaying to same sign di-electron. We did not attempt to evaluate the bounds on the DM from LHC search as the one from collider is weaker than the one obtained from direct search. 

To summarise the model is well motivated and has a rich phenomenology. It can be distinguished from the usual Higgs portal DM models from direct search prospect and with reduced SM coupling can easily survive LHC bounds. The model predicts signatures in leptonic final states at LHC through the productions of charged scalars. If signals in dilepton, four lepton and signals with higher leptonic multiplicity is seen with higher luminosity data, will indicate towards the existence of such framework.

\begin{acknowledgments}

The work of SN was supported in part by the United State Department of Energy Grant Number DE-SC 0016013. The work of SB is supported by DST-INSPIRE grant no. PHY/P/SUB/01 at IIT Guwahati. 
 
\end{acknowledgments}

\end{document}